# Radiated Energy Calculation in FELWI

I.V. Dovgan*

Department of Physics, Moscow State Pedagogical University, Moscow 119992, Russia.

K. B. Oganesyan

Alikhanyan National Science Lab, Yerevan Physics Institute, Yerevan, Armenia.

**Abstract**

The equations of particle motion in the FELWI are derived using Hamiltonian formalism. In small signal regime the uncoupled one dimensional phase equation is derived in the form of pendulum equation. For the practical estimations the same equation along with the equation of particle energy change are solved using perturbation theory and the expressions for gain in FEL regime and particle angle dependence of energy at the exit of first undulator are obtained. Results for gain, particle phase and energy change depending on device and beam parameters are presented.

1.  Introduction

Usually free-electron lasers (FELs) [1, 2] use the kinetic energy of relativistic electrons moving through a spatially modulated magnetic field (wiggler) to produce coherent radiation. The frequency of radiation is determined by the energy of electrons, the spatial period of magnetic field and the magnetic field strength of the wiggler. This permits tuning a FEL in a wide range unlike atomic or molecular lasers. There are numerous publications devoted to FELs based on undulators and strophotrons [8-50] and references therein. However for purposes of achieving a short-wavelength region of generation there are important possible limitations of the FEL gain.

   The idea of inversionless FEL or FELWI (FEL without inversion) was  discussed in  [3 -7].  More specifically, the idea of FELWI is based on a two wiggler scheme with a specially organized dispersion region between the wigglers. In principle, the two-wiggler scheme is widely used in normal FELs. This scheme (often referred to as an optical clystron) is known to provide a somewhat higher gain with narrower amplification band than in a single-wiggler FEL but it does not provide conditions for amplification without inversions.

*dovganirv@gmail.com

## 2. Equations of motion

Beginning with the equations of particle motion in the undulator field + electromagnetic field of laser governed by Hamiltonian

$$H = c\sqrt{p - (eA)^2 + m^2 c^2} \qquad (1)$$

where the combined vector potential reads

$$A = 2\hat{y}\left[A_w \cos(-v_w t - k_w z) + A_L \cos(-v_L t - k_L z \cos\theta + k_L x \sin(\theta + \phi))\right] \qquad (2)$$

one can derive the particle phase and energy change equations in the form:

$$\begin{cases} \dot{\gamma} = N \sin\psi \\ \ddot{\psi} = \dfrac{P}{N} \sin\psi \end{cases} \qquad (3)$$

where the phase $\psi = \Delta v t + q_z z + q_x x + \phi$ is expressed in terms of momentum $q_x = k_L \sin\theta$ $q_z = k_L \cos\theta + k_w$; and energy transfer $\Delta v = v_L - v_w$ during each act of photon emission.

Introducing small deviations from particle injected velocity and longitudinal coordinate $z = V_i t + \delta z$; $V_z = V_i + \delta V_z$ in case of particle energy close to resonance energy one can present the phase via the detuning $\Omega = q_z V_{zi} - \delta v$ as following

$$\psi = \Omega t + q_z \delta z + q_x x + \phi \qquad (4)$$

In above equations the parameters are read as:

$$N = \frac{e^2 2 A_w A_L \Delta v}{m^2 c^2 \gamma}; \quad P = \frac{N^2 c^2}{\gamma_r^3 \Delta v}(\gamma_r^2 q_x^2 + q_z^2) \qquad (5)$$

In deriving of second equation of (3) it was assumed that in first undulator the change of laser intensity is negligible: $A_L \approx const$.

## 3. Solution of equations of motion

The set of equations (3) contains two uncoupled equations. This simplifies the solution of this system, and we begin with the second equation, which has the analytical type of pendulum ordinary differential equation.

For the future simplicity let rewrite this equation in the form

$$\ddot{\psi} = \omega^2 \sin\psi \qquad (6)$$

with the initial conditions $\psi(0) = \phi$; $\dot{\psi}(0) = \Omega$.

In (6) we have introduced the oscillation frequency

$$\omega^2 = \frac{2K^2 \xi (4\pi v_w)^2}{1 + K^2/2 + \gamma^2(\alpha - \theta)} \qquad (7)$$

In (7) $K = \frac{eA_w}{mc} \equiv \frac{eB}{mck_w}$, is the undulator parameter, $\xi = \frac{A_L}{A_w}$ is the ratio of laser field dimensionless potential to the undulator field dimensionless potential, $v_w = c/\lambda_w$ with - $\lambda_w$ period of undulator, $\alpha$ - the angle between the particle initial velocity and undulator axis, and $\theta$ is the angle of laser propagation in respect to undulator axis. From (7) one can make the estimation $\omega \approx v_w$.

After the integration by quadrature within the limits $[0,t]$, we obtain

$$\int_\phi^\psi \frac{d\psi}{\sqrt{\Omega^2 + 2\omega^2 \cos\phi - 2\omega^2 \cos\psi}} = \pm t \qquad (8)$$

Taking in l.h.s of (8) as the difference of integrals in limits $[0,\psi]$ and $[0,\phi]$ we ends to elliptic integrals of first kind

$$u = \int_0^\psi \frac{d\psi}{\sqrt{a - b\cos\psi}} = \frac{2}{\sqrt{a+b}} F(\varphi, k) \qquad (9)$$

with the module

$$k^2 = \frac{4\omega^2 K^2 \xi (4\pi v_w)^2}{4\omega^2 + \Omega^2 - 2\omega^2(1 - \cos\phi)} \qquad (10)$$

and the argument

$$\varphi = a c r \sin\sqrt{\frac{(a+b)(1-\cos\psi)}{2(a-b\cos\psi)}}; \quad a = \Omega^2 + 2\omega^2 \cos\phi; \quad b = 2\omega^2 \qquad (11)$$

in case of phase interval $0 \le \psi \le \pi$; and

$$u = \int_0^\psi \frac{d\psi}{\sqrt{a - b\cos\psi}} = \frac{2}{\sqrt{a+b}} F(\psi/2, k) \qquad (12)$$

with the argument

$$\varphi = \psi/2 \qquad (13)$$

in case of phase interval $\pi \leq \psi \leq 2\pi$.

From the condition $k^2 < 1$ one obtains the limitation on the initial phase:

$$-\frac{\Omega}{2\omega} < \sin\phi/2 < \frac{\Omega}{2\omega} \qquad (14)$$

Taking into account that $\Omega/\omega \propto \gamma^2$ it is obvious the (13) easily fulfilled in energy region greater than few keV. From (9) and (12) we can express the particle phase via the Jacobi elliptical function

$$\sin\psi = snu = sn\left(u_0 + \sqrt{\frac{a+1}{2}}\omega t\right) \qquad (15)$$

with the corresponding values of initial phase defined from (11) and (13).

Lets turn back to set (3). For the first equation we can write down

$$\frac{d\gamma}{dt} = \frac{N^2}{P}\frac{d^2\psi}{dt^2} \qquad (16)$$

from which one can obtain

$$\Delta\gamma = \frac{N^2}{P}(\dot\psi - \dot\psi_0) \qquad (17)$$

or finally

$$\frac{\Delta\gamma}{\gamma} = \frac{\eta}{1+\eta} \qquad (18)$$

where we have introduced the parameter

$$\Delta\gamma = \frac{N^2}{P\gamma_0}(\dot{\psi} - \dot{\psi}_0) = \frac{\gamma^2 \Delta v}{c^2(\gamma^2 q_x^2 + q_z^2)}(\dot{\psi} - \Omega) \qquad (19)$$

In simulation of phase variation in (19) one may use the relation

$$\frac{d\sin\psi}{dt} = \cos\psi\,\dot{\psi} = \frac{d\,snu}{du} \times \dot{u} = cnu \times \Delta(\psi) \times \dot{u} \qquad (20)$$

Taking into account that in small signal regime the gain $G \propto \Delta\gamma/\gamma$, one can estimate the intensity of wave enhancement using above derived formulae, as well as the angle of particle in respect to undulator axis at the exit of the undulator defined as

$$\Delta\alpha_1 = \Delta\gamma_1 \frac{\theta - \alpha}{\gamma} \qquad (21)$$

## 4. Approximate solution

For the practical analytical estimations one can obtain the solution of (3) using perturbation series technique. In the zero order of perturbation one can ignore small values in the phase expression and rewrite the first equation of (3) in the form

$$\frac{d\gamma_1^0}{dt} = N\sin(\Omega t + \phi) \qquad (22)$$

In this order the average gain is zero. In the first order for the coordinates and energy change one can write

$$\Delta z^1 = \frac{q_z c^2}{\Delta v \gamma_r^3} \int_0^t \Delta\gamma_1^0(t')\,dt' \qquad (23)$$

$$x^1 = \frac{q_x c^2}{\Delta v \gamma_r} \int_0^t \Delta\gamma_1^0(t')\,dt' \qquad (24)$$

$$\frac{d\gamma_1^1}{dt} = \frac{P}{N}\cos(\Omega t + \phi) \times \int_0^t \Delta\gamma_1^0(t')\,dt' \qquad (25)$$

From (22) after integration one obtains

$$\gamma_1^0(t) = -\frac{N}{\Omega}\cos(\Omega t + \phi) \tag{26}$$

After substitution of (26) into the (25) for the energy increase at the exit of first undulator in first order of perturbations one obtains

$$\Delta\gamma_1^1 = NT\sin\left(\frac{\Omega T}{2} + \phi\right) \times \frac{\sin(\Omega T/2)}{\Omega T/2} \tag{27}$$

Where $T \approx L_w/V_r$ is the time of flight of electron through first undulator. The formula (27) repeats that one given in [7].

## 5. Conclusion

With the aim of completeness as well as for verification purposes the equations of particle motion in the FELWI are derived using Hamiltonian formalism. In small signal regime the uncoupled one dimensional phase equation is derived in the form of pendulum equation.

The pendulum equation for phase oscillations is solved exactly in the Jacobi elliptic functions.

For the practical estimations the same equation along with the equation of particle energy change are solved using perturbation theory and the expressions for gain in FEL regime and particle angle dependence of energy at the exit of first undulator are obtained. Obtained results are compared with results of different authors.